\documentclass{emulateapj}

\usepackage{ltgtsim}
\usepackage{epsfig}

\newcommand{\msun}{M_{\odot}}
\newcommand{\rsun}{R_{\odot}}
\newcommand{\lsun}{L_{\odot}}
\newcommand{\mstar}{M_{*}}
\newcommand{\rstar}{R_{*}}
\newcommand{\lstar}{L_{*}}
\newcommand{\tstar}{T_{*}}
\newcommand{\mdots}{\dot{M}_{*}}

\newcommand{\rc}{R_{\rm cen}}
\newcommand{\rcore}{R_{\rm core}}
\newcommand{\rd}{R_{\rm dust}}
\newcommand{\tdust}{T_{\rm dust}}
\newcommand{\dgr}{^{\circ}}
\newcommand{\rgr}{r_{\rm gr}}
\newcommand{\rhogr}{\rho_{\rm gr}}

\begin{document}

\title{How Protostellar Outflows Help Massive Stars Form}

\author{Mark R. Krumholz}
\affil{Physics Department, University of California, Berkeley,
Berkeley, CA 94720}
\email{krumholz@astron.berkeley.edu}

\author{Christopher F. McKee}
\affil{Departments of Physics and Astronomy, University of California,
Berkeley, Berkeley, CA 94720}
\email{cmckee@astron.berkeley.edu}

\author{Richard I. Klein}
\affil{Astronomy Department, University of California, Berkeley,
Berkeley, CA 94720, and Lawrence Livermore National Laboratory,
P.O. Box 808, L-23, Livermore, CA 94550}
\email{klein@astron.berkeley.edu}

\begin{abstract}
We consider the effects of an outflow on radiation escaping from the
infalling envelope around a massive protostar. Using numerical
radiative transfer calculations, we show that outflows with properties
comparable to those observed around massive stars lead to significant
anisotropy in the stellar radiation field, which greatly reduces the
radiation pressure experienced by gas in the infalling envelope. This
means that radiation pressure is a much less significant
barrier to massive star formation than has previously been thought.
\end{abstract}


\section{Introduction}

\label{intro}

Stars with masses $\gtsim 20\msun$ have short Kelvin times that enable
them to reach the main sequence while they are still accreting
\citep{shu87}. The resulting nuclear burning leads to a huge
luminosity, which produces a correspondingly large radiation pressure
force on dust grains suspended in the incoming gas. This force can
exceed the star's gravitational pull, possibly halting accretion and
setting an upper limit on the star's final mass. Early spherically
symmetric calculations suggested that this phenomenon sets an upper
limit on stellar masses of $\sim 20-40\msun$, \citep{kahn74,
wolfire87} for typical galactic metallicities. More recent
non-spherical calculations have loosened that constraint by
considering the role of accretion disks \citep{nakano89, nakano95,
jijina96}. Disks reduce the effects of radiation pressure by
concentrating the incoming matter into a smaller solid angle,
increasing its ram pressure. They also absorb stellar radiation in a
thin ring and re-radiate it isotropically, casting a shadow of reduced
radiation pressure. Even with a disk, however,
radiation pressure can still be a significant barrier to
accretion. \cite{jijina96} were able to form a 50 $\msun$ star only 
if it had a thin accretion disk with a radius $\sim 4000$
AU. Simulations by \citet{yorke02} in 2D found limiting masses of
$\sim 40\msun$ before radiation pressure reversed the inflow. However,
observations show that considerably more massive stars exist, and
their formation mechanism remains uncertain.

Recent observations of massive protostars have added
an element to this picture. \citet{beuther02a,beuther02b},
\citet{beuther03} and \citet{beuther04} report interferometric
measurements
showing outflows from massive protostars with collimation
factors of $\sim 2$ up to $\sim 10$. They conclude that high and low
mass outflows have similar collimation factors, typically
$2-5$ \citep{bachiller96}. (The collimation
factor is the ratio of the outflow's length to its width.)
\citet{richer00} show that the momentum of CO outflows
driven by massive stars scales with the bolometric luminosity of the
source in the same manner as for low mass stars. From these
observations, the natural conclusion is that low and high mass stellar
outflows have a common driving mechanism and similar morphologies.

Previous theoretical work on massive star formation did not include
the effects of outflows, and therefore assumed that the stellar
radiation was either isotropic (e.g. Jijina and Adams 1996) or had only
those anisotropies induced by the presence of a disk
(e.g. Yorke and Sonnhalter 2002). In this Letter, we
calculate radiative transfer through dense envelopes accreting
onto massive protostars, and we study the effects of outflow cavities in
the envelopes on the radiation field. In \S~\ref{models} we present
our models for radiative transfer and for the protostellar
environment. We then present the results of our calculations in
\S~\ref{results} and discuss conclusions in \S~\ref{conclusions}.

\section{Models}

\label{models}

\subsection{Radiative Transfer Methodology}

We used the Monte Carlo/diffusion radiative transfer code
written by \citet{whitney03a,whitney03b} to find the
temperature distribution of gas in a circumstellar envelope. The
opacity comes from dust grains, which are thermally well-coupled to
the gas in the high density envelope \citep{spitzer78}. 
The grain size distribution depends on the location in the flow. The
stellar spectrum is a Kurucz model atmosphere.
Rather than discuss the code, grain, and stellar
spectral models here, we refer readers to \citet{whitney03a}, and to
\S~\ref{outflow} for a discussion of opacity in the outflow.We used
$10^7$ photons, sufficient to produce high signal-to-noise spectra.

To calculate the radiation force on the gas, we begin with the temperature
distribution determined by the Monte Carlo code. From a given
position, we pick ray directions based on the
HEALPix scheme \citep{gorski99}. Along each ray, we solve the transfer
equation for emission and absorption by dust grains to obtain
$I_{\nu}(\mathbf{n})$, the specific intensity coming from
direction $\mathbf{n}$. We also perform this calculation along a ray
to the star itself. Once we have found the intensity along each ray,
we compute the flux
\begin{equation}
\mathbf{F}_{\nu} = \int I_{\nu}(\mathbf{n}) \mathbf{n} \; d\Omega
\approx \frac{4\pi}{N_{\rm rays}} \sum_{k=1}^{N_{\rm rays}} I_{\nu}(\mathbf{n}_k)
\mathbf{n}_k,
\end{equation}
where $N_{\rm rays}$ is the number
of rays. We increase $N_{\rm rays}$ until the change in flux is
less than 2.5\% between iterations. We then integrate over frequency
to obtain the radiation force per unit mass, $\mathbf{f}_{\rm rad} =
c^{-1} \int \kappa_{\nu} \mathbf{F}_{\nu} \; d\nu$. 

We ignore scattering when calculating the force (but not the
temperature). Scattering of IR photons is negligible.
UV and visible photons can scatter
significantly, but for the fiducial envelope and star we use,
photons at the peak of the stellar spectrum are all reprocessed into
IR in the inner few AU of the envelope. (In the disk the
distance is vastly smaller.) Outside this layer, neglecting scattering
does not change the radiation pressure force. For the same
reason, the choice of stellar spectrum model is relevant only in the inner
few AU of the envelope.

\subsection{Star, Disk, and Envelope Properties}

\label{envelope}

To study the effects of outflow cavities, we choose a single fiducial
model for the star, disk, and envelope, and vary the properties of the
outflow. We place an $\mstar=50$ $\msun$ zero-age main sequence star in
a 50 $\msun$ envelope. The ZAMS models of
\citet{tout96}, predict a radius, surface temperature, and luminosity of
$\rstar = 10.8$ $\rsun$, $\tstar=4.3\times 10^4$ K, and $\lstar=3.5\times
10^5$ $\lsun$. \citet{mckee03} predict a formation time of $\sim 10^5$ yr
for such a star, so we adopt an accretion rate of
$\mdots=5\times 10^{-4}$ $\msun$ yr$^{-1}$. The accretion luminosity
is $L_{\rm acc} \approx G \mstar \mdots/\rstar = 7.1\times
10^{4}\lsun$, negligible in comparison to the central star.

Although the cores that form massive stars are probably
turbulent \citep{mckee03}, for simplicity we adopt a simple
rotationally flattened density distribution given by \citep{ulrich76,
terebey84} 
\begin{eqnarray}
\label{TSC1}
\rho & = & -\frac{\mdots}{4\pi r^2 u_r}
\left[1+2\frac{\rc}{r}P_2(\cos\theta_0)\right]^{-1} \\
u_r & = & -\left(\frac{2 G \mstar}{r}\right)^{1/2}
\left(1+\frac{\cos\theta}{\cos\theta_0}\right)^{1/2} \\
\label{TSC2}
\frac{\rc}{r} & = & \frac{\cos\theta_0 - \cos\theta}
{\sin^2\theta_0 \cos\theta_0}.
\end{eqnarray}
Here $P_2$ is the Legendre polynomial, $r$ and $\theta$ give the
position in the envelope in spherical coordinates, and $\rc$ is the
centrifugal radius of the flow. The gas at position $(r,\theta)$
was at angle $\theta_0$ when it began falling toward the star, where
$\theta_0$ is given implicitly by (\ref{TSC2}). The centrifugal radius
$\rc$ is determined by the angular momentum of material arriving at
the star. When the last material in the core accretes, it is roughly
$\rc=\beta \rcore$, where $\beta$ is the ratio of the core's
rotational kinetic energy to gravitational binding energy and for most
cores $\beta\sim 0.02$ \citep{goodman93}. Since our
core is half accreted, we take $\rc\approx \beta
\rcore/2$.  We therefore adopt
values of $\rcore = 0.22$ pc and $\rc=400$ AU, which give the
correct mass in the envelope and satisfy  $\rc\approx \beta
\rcore/2$. 

In addition to the envelope, our star has a disk of radius
$\rc$. Observations of disks around massive stars are limited, and we
therefore adopt a disk mass of $M_d=\mstar/10=5\msun$ based on the
consideration that a disk more massive of order 10\% of the central
object is likely to be subject to gravitational instabilities
\citep{shu90} that will cause matter to accrete until marginal
stability is restored. We model the disk surface density and scale
height as power laws $\Sigma\propto r^{-2.25}$,
$h=0.035\,\left(r/\rstar\right)$ AU. We base the choice $h\propto r$
on images from simulations of massive star formation (e.g. Yorke \&
Sonnhalter 2002) which show disks with little or no flaring.

\subsection{Outflow Properties}

\label{outflow}

We can partially describe the outflow cavity from
observations. As discussed in \S~\ref{intro}, high mass
outflows have collimation factors from $2-5$,
similar to low mass stars. This
corresponds to half-angles from $0-15\dgr$,
including the uncertain
inclination. We therefore test
opening angles of $\theta_o = 5\dgr$, $10\dgr$ and $15\dgr$.
Unfortunately, observations constrain only the asymptotic
opening angle when the outflow is far
from its parent core. Cavity walls from nearby low mass
sources seen in near-IR are generally curved when viewed on
sufficiently small scales (e.g. Padgett et al 1999 and
Reipurth et al 2000). This
curvature is a natural result of outflow collimation by the
rotationally-flattened envelope \citep{wilkin98, wilkin03}. For
massive protostars that are still embedded in dense envelopes, near-IR
observations are impossible due to dust extinction. Millimeter
observations, even with interferometers, are unable to probe length
scales comparable to the size of the protostellar disk, at which
we expect the strongest collimation and curvature. Following
\citet{whitney03b}, therefore, we parameterize the
uncertain shape of the cavity as $z=a \varpi^b$, where $z$ is the
vertical distance from the star, $\varpi$ is the distance from the
outflow axis, and $a=\rcore^{1-b} \cos\theta_o /\sin^b\theta_o$ is a
constant chosen to give an opening angle of $\theta_o$ at the edge of
the core. We try values of $b=1.25$, 1.5, and 2.0, coupled with a
fixed wind opening angle $\theta_o=10\dgr$, to study the effects of
variations in cavity shape. As a baseline, we also study a case with
no outflow cavity.

Some gas at the outflow base will be ionized
by the star's UV flux (Tan \& McKee, in preparation). Whether
the ionized gas remains
near the star depends on the outflow's
structure at its base, where the density is highest and recombinations
are fastest. If the outflow is ionized, it will be too hot to contain
dust grains, and its opacity will come primarily from resonant
scattering by metal ions \citep{castor75}. This is a
complex subject beyond the scope of this Letter. We
simply note that the opacity produced by this scattering must be far
smaller than that produced by dust grains in the envelope. In the case
that the outflow cavity is ionized, we may therefore set the
opacity in the cavity to zero.


\begin{figure}
\epsfig{file=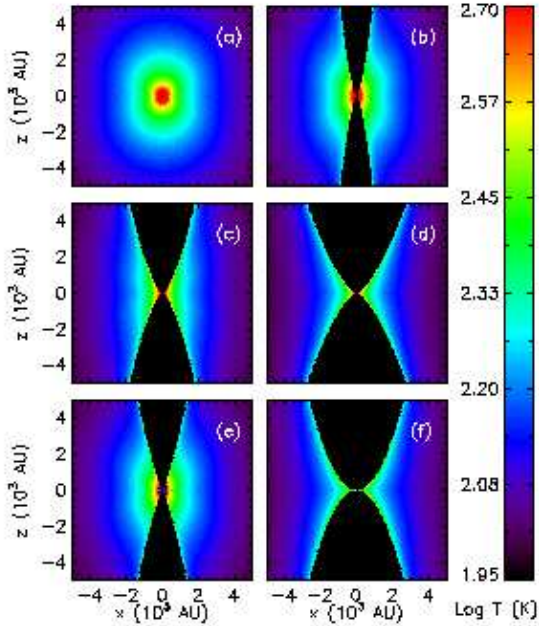}
\caption{\label{tempmap}
The color maps show the gas temperatures for each of our models. The
models are (a) no wind cavity, (b) $\theta_o=5\dgr$, $b=1.5$, (c)
$\theta_o=10\dgr$, $b=1.5$, (d) $\theta_o=15\dgr$, $b=1.5$, (e)
$\theta_o=10\dgr$, $b=1.25$, and (f) $\theta_o=10\dgr$, $b=2.0$. The
red dots inside the cavity in panel (c) are the result of a minor
code bug.
}
\end{figure}

If the outflow is neutral, grains will re-form, growing in radius
$\rgr$ at rate (cf. Hoyle 1946)
\begin{equation}
\label{graingrowth1}
\dot{r}_{\rm gr} = \alpha v_t \rho/(4 \rhogr),
\end{equation}
where $\alpha$ is the mass fraction of the element of which the grain
is composed, $v_t$ is the thermal velocity of gaseous atoms of that
element, $\rho$ is the gas density, and $\rhogr$ is the grain density.
Growth is fastest for carbon grains, since carbon is
the most abundant refractory metal in the galactic ISM and has a large
thermal
velocity due to its low atomic mass. \citet{sofia01} estimate its
mass fraction to be $\alpha_C \approx 3\times
10^{-3}$. Grains can only condense below the dust destruction
temperature of $\tdust\approx 1600$ K, so $v_t\ltsim 1$ km
s$^{-1}$. The density of carbon grains is $\sim 1$ g cm$^{-3}$. For
our adopted cavity shape, the wind density at a distance $z$ from the
equatorial plane must be roughly $
f_w \mdots/[2 \pi v_w \left(z/a\right)^{2/b}]$,
where $v_w$ is the wind velocity (taken to be constant) and
$f_w$ is the fraction of mass reaching the star that is ejected into
the wind. We take $f_w=0.2$ and $v_w=v_K=(G\mstar/\rstar)^{1/2}\approx
1000$ km s$^{-1}$ \citep{richer00}, where $v_K$ is the Keplerian
velocity at the stellar surface.

\begin{figure}
\epsfig{file=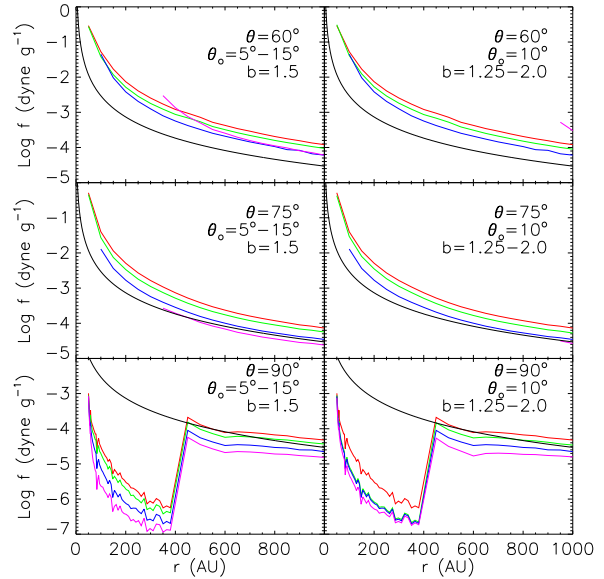}
\caption{\label{angplot}
The plots show the force per unit mass due to radiation and gravity
(\textit{black line}), at angles of
$\theta=60\dgr$, $75\dgr$, and
$90\dgr$ from the polar axis. The plots begin at the minimum value of
$r$ that is inside the infalling envelope. The three left panels show
radiation with no cavity (\textit{red
line}), $\theta_o=5\dgr$ (\textit{green line}), $\theta_o=10\dgr$
(\textit{blue line}), and $\theta_o=15\dgr$ (\textit{purple
line}), all with $b=1.5$. The three right panels show radiation with no
cavity, $b=1.25$ (\textit{green line}), $b=1.5$ (\textit{blue
line}), and $b=2.0$ (\textit{purple line}), all with
$\theta_o=10\dgr$.
}
\end{figure}

The smallest distance at which grains can form is the dust destruction
radius, which we calculate to be $\rd\approx 48$ AU for our fiducial
model. We may therefore integrate (\ref{graingrowth1})
from $\rd$ to $\rcore$ to obtain the maximum size that grains can
attain before escaping the core. The largest grain sizes occur for
$\theta=5\dgr$ and $b=1.5$, which gives
$\rgr \leq 1.1\times 10^{-4}$ $\mu$m -- in effect, grains cannot
grow past the size of molecules. In contrast, the typical grain
sizes in the envelope are $0.1-0.2$ $\mu$m. Stellar and envelope
radiation has wavelengths $\lambda \gtsim 2\pi \rgr$, so grain opacity
scales as $\kappa\propto \rgr$. Thus, even if all the C goes into
grains, the opacity in the outflow is
smaller than that in the envelope by at least a factor of $\sim
800$. If we included cooling of the outflow and evaluated the grain size
at radii smaller than $\rcore$, the reduction would be even larger. We
conclude, therefore, that dust opacity in a neutral outflow is
negligible in comparison to the opacity in the infalling
envelope. Since for either a neutral or an ionized outflow the opacity
is negligible, we set the opacity in the outflow cavity to zero in our
calculations.

\section{Results and Discussion}

\label{results}

We show the gas temperature
distribution for each of our models in Figure \ref{tempmap}. Although
our core extends to 0.22 pc, we concentrate on the
inner few thousand AU, where the radiation pressure force is
strongest. The plot shows
that, in models with a cavity, the gas is hotter on the inside edge of
the outflow cavity but is cooler elsewhere. Since, due to rotational
flattening, most of the gas
is attempting to accrete close to the equatorial plane, this means
that the bulk of the accreting gas is cooler in runs with an
outflow. The effect is stronger for outflow cavities with
larger opening angles, and for cavities that are wider at their base.

We show the radiation pressure force versus radius at angles of
$\theta=60\dgr$, $75\dgr$, and $90\dgr$ from the pole in Figure
\ref{angplot}. Consistent
with the reduced temperature, the radiation pressure force is also
substantially smaller in runs with an outflow cavity. For our
``intermediate'' model with $\theta_o=10\dgr$, $b=1.5$, the radiation
pressure force is smaller than in the run with no cavity by a factor
of up to 4.6. For the other runs, the peak reductions in radiation
force were factor of 1.7 ($\theta_o=5\dgr$, $b=1.5$), 14.4
($\theta_o=15\dgr$, $b=1.5$), 4.2 ($\theta_o=10\dgr$, $b=1.25$), and
7.2 ($\theta_o=10\dgr$, $b=2.0$). Thus, the amount by which the wind
cavity reduces radiation pressure shows moderate dependence on the
wind curvature and opening angle, but for only
one combination of parameters was the reduction less
than a factor of 4. Comparing radiation and
gravitational forces, it is clear that this reduction can mean the
difference between accretion halting or continuing.

To check how the results depend on properties of the envelope, which has a
comparatively low surface density compared to most high-mass cores, we
re-ran the intermediate and no cavity cases with
identical $\rc$ and $\rcore$, but a 100 $\msun$ envelope. The case
with no wind cavity showed little change in force with envelope mass.
With a cavity, the increased envelope mass decreased the radiation
pressure force at equatorial angles so that radiation was stronger
than gravity only at $\theta \ltsim 60\dgr$, increasing the fraction
of solid angle through which accretion could occur.
Thus, our results likely represent a \textit{lower
limit} on the reduction in radiation pressure force that outflow
cavities actually produce.

To determine how the effect of the cavity compares
to that produced by the disk and the rotationally-flattened envelope
alone, we also considered a spherical envelope, with no
cavity, no disk, and $\rc=0$ in (\ref{TSC1})-(\ref{TSC2}). We compared
this to the run with our fiducial disk and envelope parameters and no
cavity. At angles $\gtsim 75\dgr$, the radiation pressure force in the
spherical case was a factor of a few higher, while at angles $\ltsim
60\dgr$ it was comparable or smaller. Thus, our intermediate wind
cavity reduces the radiation pressure force relative to the
disk-and-envelope only case by about the same amount that
the disk-and-envelope case reduces it relative to purely spherical.
Collimation of the radiation field by the disk and
envelope and collimation by the outflow
cavity are about equally important, and reinforce each
other. Together, they reduce the radiation pressure force in the
intermediate cavity case by a factor of $\sim 10$ relative what one would
find for an isotropic radiation field.

In all our tests, the degree of radiation collimation
is roughly consistent with expectations. Tan \& McKee (in preparation) 
show that the fraction of radiation escaping an envelope through a
path of optical density $\tau$ is proportional to
$(1+\tau)^{-1}$. In our case, the flux fraction escaping from the core
at angles between $\theta$ and $\theta+d\theta$ should roughly satisfy
\begin{equation}
\label{fluxfrac}
dF(\theta) \propto \left[1+\tau(\theta)\right]^{-1}
d\left[\cos(\theta)\right],
\end{equation}
where $\tau(\theta)$ is the optical depth at angle $\theta$. 
There is considerably uncertainty in applying this to our problem
because $\tau(\theta)$ is frequency-dependent. However,
if we use the Rosseland mean opacity at
$\tdust=1600$ K, then at most angles (\ref{fluxfrac}) predicts
roughly the correct flux fraction.

\section{Conclusion}

\label{conclusions}

We have shown that an outflow can substantially change the
radiation field, and radiation pressure, around a massive protostar.
The outflow cavity provides an optically thin channel through
which radiation can escape, significantly
reducing the radiation pressure. With no wind
cavity, in our fiducial model the radiation pressure
force is stronger than gravity essentially everywhere except inside
the accretion disk, and it is therefore likely that accretion would be
halted. In our intermediate model, outside the centrifugal radius
radiation is weaker than gravity over about $\pi$ sr, providing a
large funnel through which accretion can continue.
The calculation we have performed here is only a proof of
principle, and we are currently performing 3-D radiation
hydrodynamic AMR simultions. Our results
strongly argue, however, that the presence of outflows provides a
mechanism for circumventing the radiation pressure limit to
protostellar accretion. Surprisingly, outflows that drive gas out of a
collapsing envelope may increase rather than decrease the size of the
final, massive star.

\acknowledgments We thank Jonathan Tan for useful discussions. This
work was supported by: NASA GSRP grant NGT 2-52278 (MRK); NSF grant
AST-0098365 (CFM); NASA ATP grant NAG 5-12042 (RIK and CFM);
and the US
Department of Energy at the Lawrence Livermore National Laboratory
under the auspices of contract W-7405-Eng-48 (RIK).

\end{document}